\documentclass[aps,twocolumn,showpacs,prl,floatfix]{revtex4}
\usepackage{graphicx}
\usepackage{epsf}
\usepackage{wrapfig}
\usepackage{epsfig}

\def\b0{{\mbox{\boldmath$0$}}}

\def\b0{{\mbox{\boldmath$0$}}}

\def \b #1{ {\bf #1}}
\newcommand{\be}{\begin{eqnarray}}
\newcommand{\ee}{\end{eqnarray}}

\def \b #1{ {\bf #1}}

\newcommand{\CM}{{\cal M}}

\def \b #1{ {\bf #1}}

     \font\tenbifull=cmmib10 scaled 1200 
     \font\tenbimed=cmmib9
     \font\tenbismall=cmmib7
       \def\bmit{\fam9 }
\mathchardef\bbkappa="7114
\mathchardef\bbrho="711A
\mathchardef\bbsigma="711B
\mathchardef\bbtau="711C
\mathchardef\bbvarrho="7125
\mathchardef\bbvarsigma="7126
\mathchardef\bbxi="7118
\def\boldkappa{{\bmit\bbkappa}}
\def\boldrho{{\bmit\bbrho}}

\begin{document}
\date{\today}
\title{On  the Interpretation of the  Processes 
  $\bf ^3He(e,e'p)^2H$  and  $\bf ^3He(e,e'p)(pn)$ at High Missing Momenta }
\author{C.Ciofi degli Atti}
\author{L.P. Kaptari}\altaffiliation{On leave from  Bogoliubov Lab.
      Theor. Phys.,141980, JINR,  Dubna, Russia}
\affiliation{Department of Physics, University of Perugia and
      Istituto Nazionale di Fisica Nucleare, Sezione di Perugia,
      Via A. Pascoli, I-06123, Italy}
\vskip 2mm

\begin{abstract}
\vskip 5mm
Using realistic three-body wave functions corresponding to
 the $AV18$ interaction, it is shown that the effects  of the Final State Interaction (FSI) in the
exclusive  processes 
$^3He(e,e^\prime p)^2H$ and  $^3He(e,e'p)(pn)$, can be successfully treated in terms of a 
Generalized Eikonal Approximation (GEA)  based
upon  the direct calculation of the  Feynman diagrams describing  the rescattering
of the struck nucleon. The relevant  role played by  the  double rescattering
 contribution at high values of the missing momentum is illustrated.

\end{abstract}
\maketitle
Recent experimental data from Jlab  on exclusive electro-disintegration of
$^3He$  \cite{jlab1,jlab2,prelim}
are at present the object of intense theoretical activity (see 
\cite{laget,claleo, rocco} and References therein).
 In Ref. \cite{claleo} (to be quoted hereafter as $I$) the 2-body  break up (2bbu)
  and 3-body break up (3bbu) channels,
 $^3He(e,e^\prime p)^2H$ and  $^3He(e,e'p)(pn)$  respectively, have been calculated within the  
 following approach:
i)initial state
correlations (ISC) have been taken care of  by the use of  the status-of-the-art
few-body wave functions~\cite{pisa} corresponding to the $AV18$
interaction \cite{av18};
ii)final state interactions (FSI) have been
treated by a  Generalized Eikonal Approximation (GEA) \cite{mark}. GEA
represents an   extended Glauber-type  approach (GA) \cite{glauber} based upon 
 the evaluation of the relevant
  Feynman diagrams that describe  the rescattering of the
   struck nucleon in the final state, in analogy with the
  Feynman  diagrammatic approach  developed for the treatment
  of elastic hadron-nucleus scattering  \cite{gribov,bertocchi}. 
  The Feynman diagrams pertaining to the process
   $^3He(e,e^\prime p)X$ (with $X$ = $^2H$ or $(pn)$) are shown in Fig. 1.
In $I$ the theoretical calculations have been compared with the preliminary Jlab data
\cite{prelim}
covering a region of missing momentum and energy with 
 $ p_m \leq\, 0.6 \, \, GeV/c$ and  $E_m\,
 \leq 100 \,MeV$. Since the  recently published data  \cite{jlab1,jlab2} 
 extend to higher values of the missing momentum  ($p_m \leq 1.0\,\, GeV/c$), which  have not been considered in $I$,
   the aim of this paper is to analyze our predictions  in the  entire range of missing
   momentum   $ p_m \leq\, 1.2 \, \, GeV/c$.  We will first of all argue  
  that the missing momentum dependence of the 2bbu channel cross section, namely its
    changes of slope,  can be  a signature
  of different orders of final state rescattering and,  secondly, 
   we will show that our fully parameter free calculation can reproduce 
   the experimental data in the entire range of missing momentum, 
    with the high missing momentum behaviour ($ p_m \geq\, 0.6 \, \, GeV/c$)  mainly
  governed by double scattering effects. 
  Although the details  of the theoretical  approach  
  can be found in $I$, it is worth briefly recalling  the basic assumptions 
  underlying  GA and GEA. The former  is based upon the following approximations:
 i) the nucleon-nucleon  (NN) scattering amplitude is
 obtained within the eikonal approximation; ii) the
nucleons of the spectator  system  $A-1$ are stationary during the multiple scattering
with  the struck
nucleon  (
{\it frozen approximation});  iii)
only perpendicular
 momentum transfer components in the NN scattering amplitude
are considered.
 In  GEA the    {\it frozen approximation} is partly removed by taking
 into account the excitation energy of the  $A-1$ system,
  resulting  in a longitudinal momentum transfer dependence of the 
   standard GA
 profile function.
 \begin{figure*}[!htp]
\centerline{
      \epsfxsize=3.2cm\epsfysize=2.5cm\epsfbox{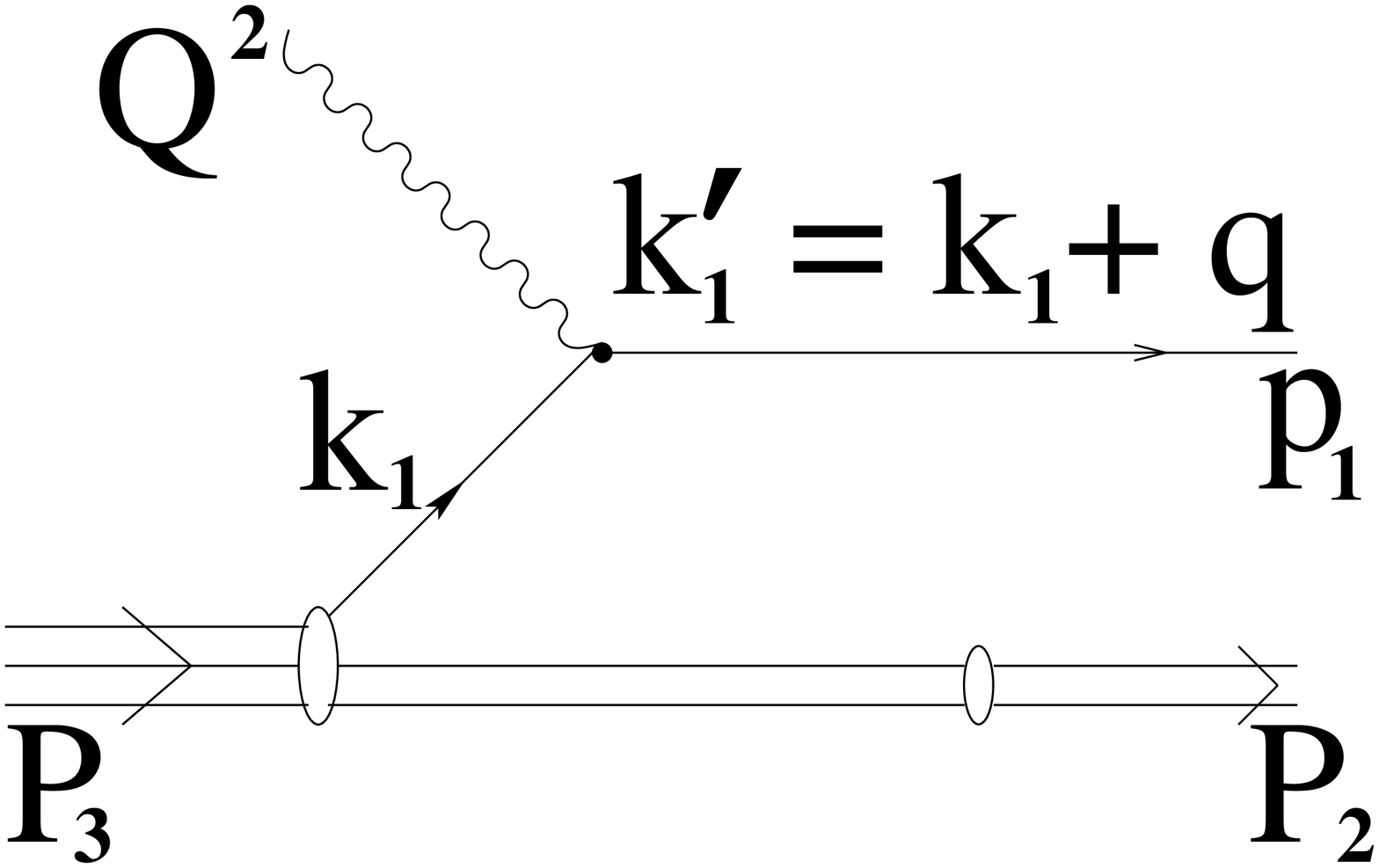}\hspace{1cm}
      \epsfxsize=3.2cm\epsfysize=2.5cm\epsfbox{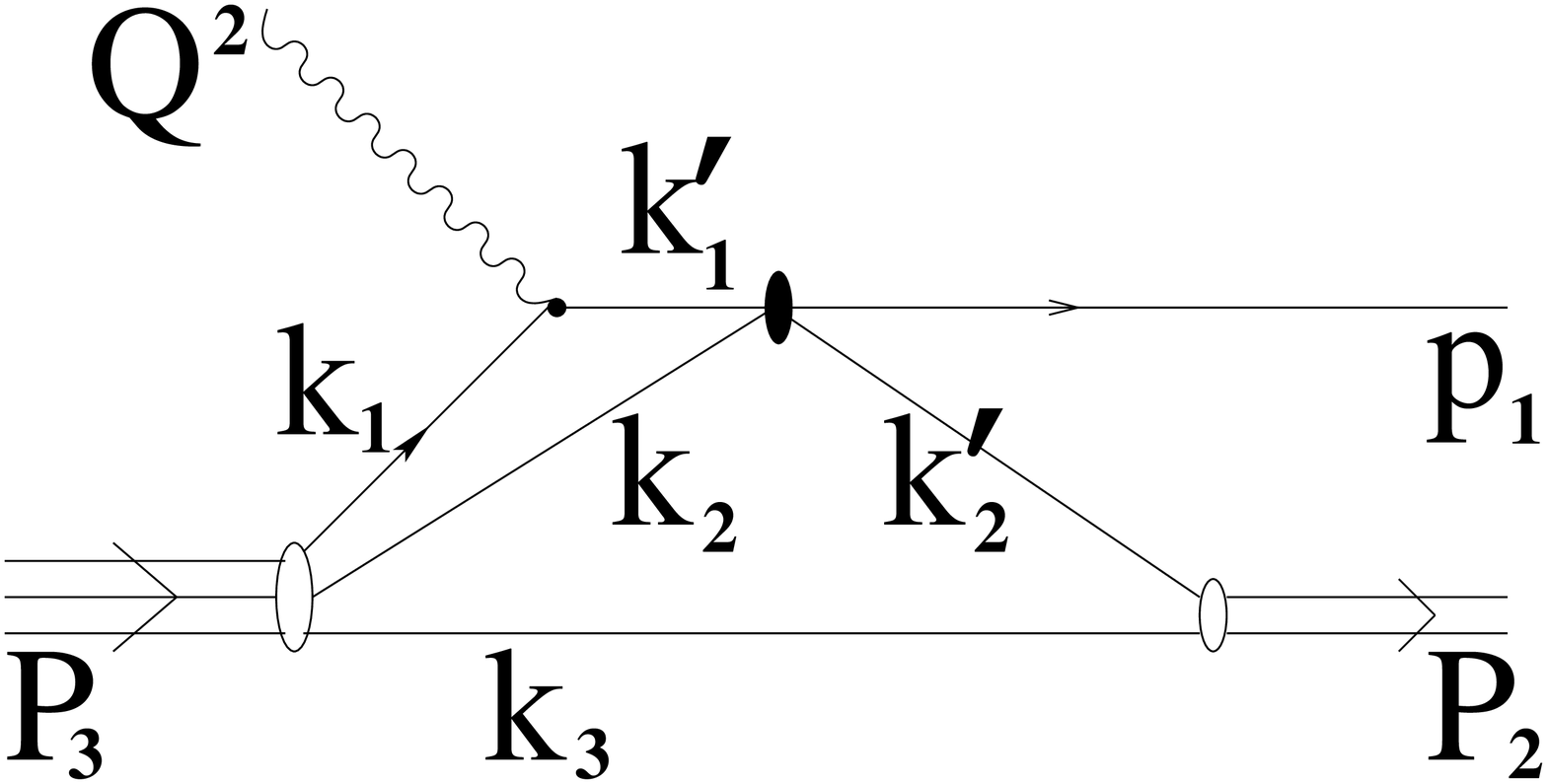}\hspace{1cm}
      \epsfxsize=4.0cm\epsfysize=2.5cm\epsfbox{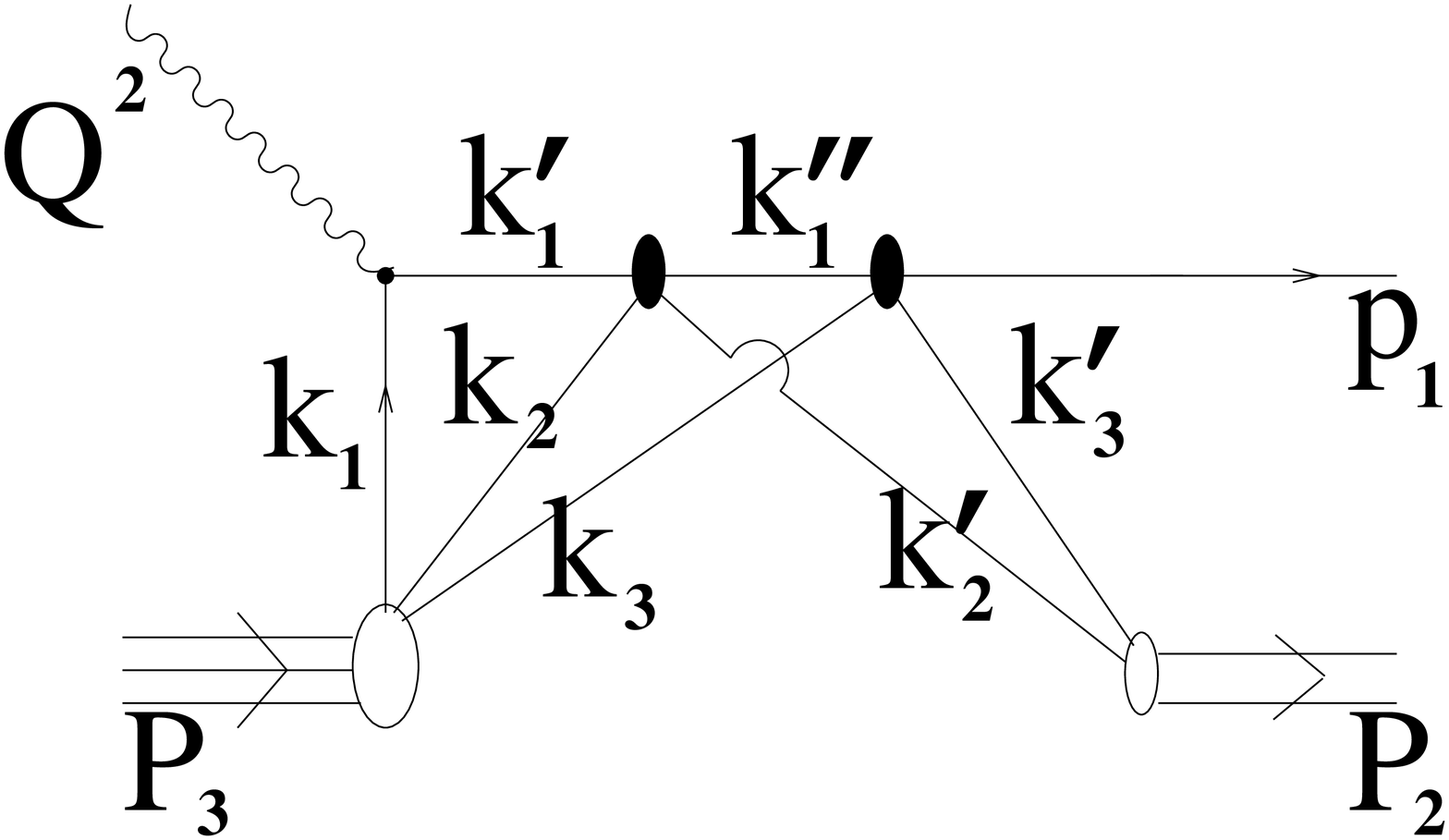}}
\centerline{\hspace{-1cm}
      \large\textbf{a)}\hspace{4.cm}
      \large\textbf{b)}\hspace{4.cm}
      \large\textbf{c)}\normalsize}
\caption{The Feynman diagrams representing the PWIA (a), the single (b),
 and double (c) rescattering in the
processes $^3He(e,e'p)^2H$ and $^3He(e,e'p)(np)$.
 In the former case the final two-nucleon state is a
deuteron with momentum ${\b P}_2$,
whereas in the latter case the final state represents two
nucleons in the continuum with momenta ${\b p}_2$ and  ${\b p}_3$, with ${\b P}_2= {\b p}_2
+ {\b p}_3$. 
 The trivial
single and double rescattering diagrams
with nucleons "2" and "3" interchanged are not drawn. The  black oval spots denote the elastic
nucleon-nucleon (NN) scattering
matrix.}
\label{Fig1}\vskip -0.1cm
\end{figure*}
 Let us omit (see $I$ for details) any discussion concerning  
 the assumptions and approximations  which, starting from the Feynman diagrams shown 
 in Fig. 1,
 allow one to obtain a calculable cross section,  and 
 let us instead write down the final expression for the latter;  
  for the processes  $^3He(e,e'p)X$,  one has
\be
&&\hspace{-0.5cm}
\frac{d^6\sigma}{dE' d\Omega' d{\b p}_m}=\nonumber\\
&&= K({Q}^2,x,{\bf p}_m)\,
 \sigma^{eN}({\bar Q}^2,{\bf p}_m)
  P_{He}^{FSI}(\b{p}_m,E_m),
 \label{crosshe}
\ee
\noindent where 
$Q^2= -q^2 = -(k-k')^2 = {\b q}^2 - q_0^2 = 4EE'\sin^2\theta/2$ is 
 the four-momentum transfer; 
$x={Q}^2/2M_N q_0$ \,\,is  the Bjorken scaling variable;
 ${\bar Q}^2$ =${\b q}^2 - {\bar q}_0^2$
\, ($\bar{q}_0 = q_0 + M_3 -
\sqrt {({\b k}_1^2 + (M_{2}^f)^2} -
\sqrt{ {{\b k}_1}^2 + M_N^2}$);   $M_3$ is the mass of  $^3He$
and $M_2^f=M_2+E_2^f$   the mass of the two-nucleon system
in the final state with  intrinsic excitation energy $E_2^f$;  
 ${\bf p}_m={\b q} - {\b p}_1 = {\b P}_2$\, and  $E_{m} = E_{min} + E_{2}^f$ 
 are  the missing momentum and  energy, respectively;  the threshold and 
  intrinsic excitation energies are 
  $E_{min}=E_3-E_2 \,\,\,(E_{min}=E_3)$\, and  $E_2^f = 0$
  \,\, ( $E_2^f={\bf t}^2/M_N$,\, ${\b t}= 
 \frac {{\b k}_2 - {\b k}_3}{2}$)\, for the 2bbu (3bbu) channel, respectively;
 $E_2\,\,(E_3)$ are  the (positive) ground-state
  energy of  $^2H\,\,(^3He)$.
The quantity $K({ Q}^2,x,{\bf p}_m)$  is  a kinematical factor,
$\sigma^{eN}({\bar Q}^2,{\bf p}_m)$ is the
 cross section
 describing electron scattering by an off-shell nucleon, and, eventually, $
 P_{He}^{FSI}({{\b p}_m},E_m)$ is the Distorted Spectral Function
containing the effects from  ISC and  FSI; within GEA, one has

\begin{widetext}
\be
P_{He}^{FSI}({{\b p}_m},E_m)\,=\, \frac {1}{(2\pi)^3} \frac{1}{2}\,\,\, \sum_{f}\,\,\,
\sum_{{\cal M}_3,\,{\cal M}_{2},\,\,s_1}
\left | \sum_{n=0}^{2} {\cal T}_3^{(n)}(\CM_3,{\cal M}_{2}, s_1;f) \right |^2
\,\delta\left( E_{m}-(E_{2}^f + E_{min})\right),
\label{pdistor3}
\ee
\end{widetext}
\noindent where
${\cal M}_3$,  ${\cal M}_2$ and $s_1$ are the magnetic quantum numbers of $^3He$, of the
two nucleon system  in the final state, and of the struck nucleon in
 the continuum, respectively; the quantity 
  ${\cal T}_3^{(n)}(\CM_3,{\cal M}_{2}, s_1;f)$, which  can be called
the {\it reduced (Lorentz index independent) amplitudes} (see e.g $I$), 
results from the evaluation of  the Feynman
diagrams of Fig. 1, with   ${\cal T}_3^{(0)}$ corresponding  to the PWIA (Diagram 1(a)),  ${\cal T}_3^{(1)}$
to the single-rescattering FSI (Diagram 1(b)), and ${\cal T}_3^{(2)}$ to the 
 double rescattering FSI
(Diagram 1(c)); the
explicit evaluation of the diagrams yields
\be
 P_{He}^{FSI}({\b p}_m,E_m) =  P_{gr}^{FSI}({\b p}_m,E_m) +
  P_{ex}^{FSI}({\b p}_m,E_m),
\label{pitotfsi}
\ee
\noindent where the first and the second terms describing  the 2bbu and the 3bbu channel,
respectively, are
\begin{widetext}
\be
\label{ngrfsi} 
\hspace{-0.3cm}P_{gr}^{FSI}({\b p}_m,E_m)&=&\frac{1}{(2 \pi)^3} \frac{1}{2}
  \sum_{\CM_3, \CM_2,s_1}
  \left | \int {\rm e}^{i\boldrho{\b p}_m}
 \chi_{\frac12 s_1}^{\dagger} \Psi_{D}^{{\CM_2}\,\dagger}(\b{r} )
  {\cal S}_{\Delta}^{FSI}(\boldrho,{\bf r})
 \Psi_{He}^{\CM_3}(\boldrho,\b{r})   d \boldrho d {\bf r} \right |^2
\delta(E_m - (E_{3}-E_2)),\\
\label{piexfsi}
\hspace{-0.3cm}P_{ex}^{FSI}({\b p}_m,E_m)&=&\frac{1}{(2 \pi)^3} \frac{1}{2}
  \hspace{-0.1cm}\sum_{{\cal M}_3, S_{23}, s_1}\hspace{-0.3cm}
       \int \frac{d^3 \b {t}}{(2\pi)^3}
  \left | \int{\rm e}^{i\boldrho{\b p}_m} \chi_{\frac12 s_1}^\dagger
 \Psi_{np}^{\b{t}\dagger}(\b{r}){\cal S}_{\Delta}^{FSI}(\boldrho,\b{r})
  \Psi_{He}^{{\cal M}_3}(\boldrho,\b{r}) d\boldrho d {\bf r}
  \right |^2\hspace{-0.2cm}
\delta \left( E_m - \frac{\b {t}^2}{M_N} - E_3 \right)
\ee
\end{widetext}
\noindent with  $\Psi_{He}^{\CM_3}(\boldrho,\b{r})$, $\Psi_{D}^{\CM_2}(\b{r} )$, and 
$\Psi_{np}^{\b{t}}(\b{r})$, being   the wave functions of  $^3He$, 
of  the deuteron, and  of the  $np$ pair in the continuum, respectively.
Here, the FSI factor   ${\cal S}_{\Delta}^{FSI}$, which   describes the single and double
 rescattering of  nucleon "1",  has the form
$
{\cal S}_{\Delta}^{FSI}(\boldrho,\b{r}) ={\cal S}_{(1)}^{FSI}(\boldrho,\b{r})+
{\cal S}_{(2)}^{FSI}(\boldrho,\b{r})
\label{totalS}
$
 \,\,with  the single and double rescattering contributions
given respectively by
\vskip -0.3cm
\begin{widetext}
\vskip -0.5cm\be
&&\hspace{-0.6cm}\label{essesingle}{\cal S}_{(1)}^{FSI}(\boldrho,\b{r})=1-\sum\limits_{i=2}^3
\theta(z_i-z_1){\rm e}^{i\Delta_z (z_i-z_1)} \Gamma (\b {b}_1-\b{b}_i)\,;\\
&&\hspace{-0.6cm}{\cal S}_{(2)}^{FSI}(\boldrho,\b{r})= \Gamma(\b {b}_1-\b{b}_2)
\Gamma(\b {b}_1-\b{b}_3)\times\nonumber\\
&&\label{essedouble}\hspace{-0.3cm}\times\left[\theta(z_2-z_1)\theta(z_3-z_2)
  {\rm e}^{-i\Delta_3(z_2-z_1)}
  {\rm e}^{-i(\Delta_3-\Delta_z)(z_3-z_1)}\,+\,
  \theta(z_3-z_1)\theta(z_2-z_3){\rm e}^{-i\Delta_2(z_3-z_1)}
      {\rm e}^{-i(\Delta_2-\Delta_z)(z_2-z_1)}\right],
\ee
\end{widetext}
\noindent where  $\Delta_i=(q_0/|{\b q}|)(E_{{\b p}_i} - E_{{\b k}_i^{'}})$ and 
$\Delta_z = ({q_0}/{|{\b q}|}) E_m$ \cite{mark}.
The usual Glauber FSI factor \cite{nikolaev}
\begin{equation}
{\cal S}_G^{FSI}(\boldrho,\b{r}) =
\prod\limits_{i=2}^{3}\ \left[ 1-\theta(z_i-z_1)\,
\Gamma({\bf b}_i-{\bf b}_1)\right ],
\label{eq20}
\end{equation}
is recovered by setting $\Delta_i = \Delta_z = 0$,
 whereas by also setting   $\Gamma = 0$, 
  the usual Spectral Function  is obtained. An inspection at 
  Eqs. (\ref{essesingle}) and (\ref{essedouble})  shows  that 
by taking into consideration the factor $\Delta_z$, the  
  {\it frozen approximation} is partly removed, resulting in  a longitudinal momentum transfer
correction term to the standard
 profile function of GA.
 Note that the conditions for the validity of the factorized form (1) for the
  cross section,
 have been discussed in detail in Ref. [5] (Appendix B), showing that if the 
 spin-flip part
 of the NN scattering amplitude is small and the momentum transfer in 
 the final state rescattering  ${|\boldkappa|} << |{\bf p}_1|$,  as it is the case
  for the sharply forward peaked NN scattering  amplitude used in Glauber-type calculations,
   then factorization should  occur to a large extent.
 Concerning the numerical results we have obtained, it is also worth mentioning 
 that we found the
 effects due to the $\Delta$'s appearing in Eqns. (6) and (7) to be of the order of few percent,
 since, as explained in Ref. [5], the experimental data we are considering are
  practically
 at perpendicular kinematics whereas  the  $\Delta$'s mainly affect the longitudinal
  momentum distributions (see also Ref. [9]).
 
 We have used Eqs.(\ref{crosshe}) and  (\ref{pitotfsi})
to calculate the cross sections of the processes 
$^3He(e,e'p)^2H$ and $^3He(e,e'p)(np)$. All calculations have been performed in the
reference frame
where the axis $z$ is directed along the momentum  of the struck nucleon ${\bf p}_1$.  
The common well known parameterization of the
profile function
$
\Gamma({\bf b}) =\sigma_{NN}^{tot}(1-i\alpha_{NN})\times
{\exp}(-{\bf b}^2/2b_0^2)/(4\pi b_0^2)
$
has been used,  where $\sigma_{NN}^{tot}$ is the total $NN$ cross section,
 $\alpha_{NN}$  the ratio of the real to imaginary part of the forward $NN$ amplitude,
 and  $b_0$ the slope of
  the differential elastic $NN$ cross section; the   energy dependent values 
 of these quantities   have been taken from Ref. \cite{said}. The  
  electron-nucleon  cross section  $\sigma_{cc1}^{eN}({\bar Q}^2,{\bf p}_m)$ is the one 
  of  Ref.
 \cite{forest}. All two- and three-body wave functions are direct
solutions of the  non relativistic Schr\"odinger equation, therefore
our calculations  are fully parameter free.
It should be stressed, in this connection,  that besides the GEA, no
further approximations have
  been made in the evaluation
  of the single and double rescattering contributions to the FSI: proper
  intrinsic coordinates have been used and the energy dependence of the profile
  function  has been taken into account in the properly chosen CM system of the
  interacting pair. 
  
\begin{figure}[!hpt]\vskip -0.5cm
\includegraphics[height=8.0cm,width=8.0cm]{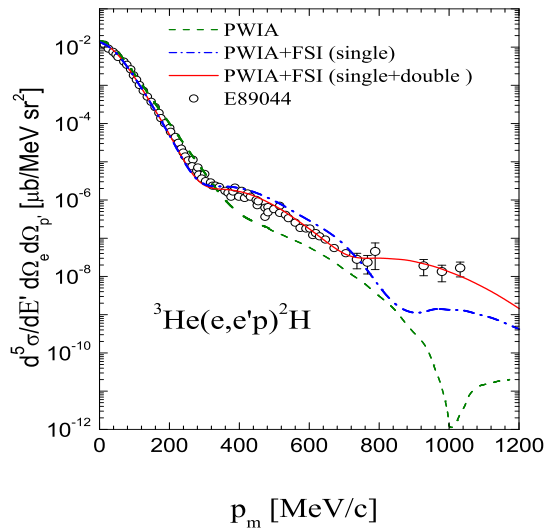}
\vskip -0.3cm\caption{The  experimental data from
JLab \cite{jlab1} ($Q^2 = 1.55\,\, (GeV/c)^2$,   $x=1$) on the 2bbu process
$^3He(e,e'p)^2H$ 
 {\it vs} the missing momentum $p_m$,   compared with
our theoretical results. The dashed line corresponds to the
PWIA  (Eq. (\ref{ngrfsi}) with ${\cal S}_{\Delta}^{FSI}=1$), the dot-dashed line includes
the FSI with single rescattering (${\cal S}_{\Delta}^{FSI} = {\cal S}_{(1)}^{FSI}$),
 and the full line
includes both single and double rescattering (${\cal S}_{\Delta}^{FSI} =
 {\cal S}_{(1)}^{FSI}$
+${\cal S}_{(2)}^{FSI}$)
     (three-body wave function from \protect\cite{pisa},  $AV18$ interaction
     \protect\cite{av18}).}\label{Fig2}\vskip -0.1cm
\end{figure}

The results of our calculation for the 2bbu channel, which  are compared with 
the experimental data in  Fig. 2, deserve the following comments:
 i) the missing momentum dependence of the experimental cross section clearly exhibits
 different slopes, that are reminiscent of the slopes  observed in elastic
 hadron-nucleus scattering at
 intermediate energies (see e.g. Ref. \cite{elastic});
ii) our calculations, as clearly illustrated in Fig. 2,  demonstrate that
 these slopes are indeed  related to multiple scattering in the final state;
iii)without any free parameter,
 a very  satisfactory agreement
 between experimental data and theoretical calculations 
 can be obtained, which means that in the energy-momentum range covered by the data,
  FSI can be described 
 within the GEA. 
\begin{figure}[!htp]\vskip -0.5cm
\epsfig{file=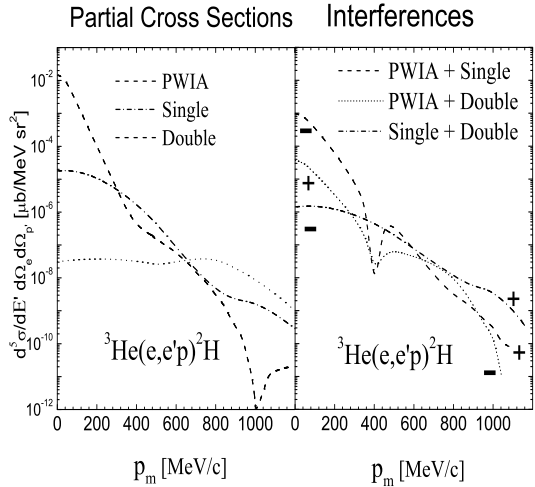,width=8cm,height=7.5cm}
\vskip -0.5cm\caption{ The partial cross sections (left) and the interference contributions
(right) in the process  $^3He(e,e^\prime p)^2H$ calculated
including  the full
FSI. The continuous curve in Fig. 2  is obtained by summing the partial cross sections
and the interference contributions with the proper sign.}
\label{Fig3}
\end{figure}
In order to better understand the role played by multiple  rescattering in the final state,
we show in Fig. 3 the separate contributions of the PWIA,  the single and double 
rescattering contributions, and the interference contributions.
It can be seen that  at $p_m \leq 0.2\,\, GeV/c$,  the cross section is mainly governed by 
the PWIA, at  $ 0.2 \leq\, p_m \,\leq \,0.6 \,\, GeV/c$, by the single rescattering FSI,
and at $  p_m \geq 0.6\,\, GeV/c$, by the  double rescattering FSI, 
with the interference terms also providing relevant contributions in specific regions.
The results of the  3bbu channel calculations are 
 shown  in Fig. 4, and they also appear to be
 in  good agreement with the experimental data \cite{jlab2},
  although a systematic underestimation
 of the latter is observed  at $p_m=820\,\, MeV/c$ and $E_{rel} \leq \,\,110\,\, MeV$.
  The origin of such a disagreement, which might be due to MEC effects [18], non factorizing contributions to
 the cross section [6], charge-interchange amplitudes, and other effects which
  are not included in our approach, is
 under investigation.
 
  \begin{figure}[!hpt]
\vskip -0.1cm\includegraphics[height=8.5cm,width=8cm]{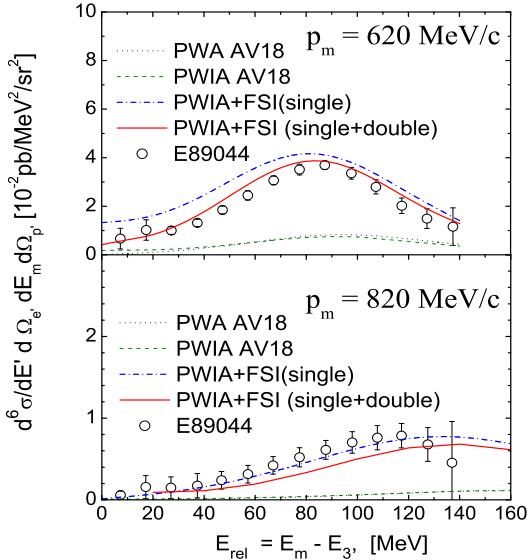}
\vskip -0.5cm\caption{The differential cross section   of  the 3bbu process $^3He(e,e'p)(np)$ 
\cite{jlab2},  plotted, for fixed values of $p_m$, {\it vs}
the excitation energy of the two-nucleon system in the continuum 
  $E_{rel}={\b {t}^2}/{M_N}$ = $E_2^f = E_m- E_{3}$. 
  The theoretical calculations correspond to Eq. (\ref{piexfsi}) and 
the meaning of the various curves
is the same as in Fig.2. The curves labelled $PWA$ do not include any FSI.}
\label{Fig4}\vskip -0.1cm
\end{figure}

To sum up, in this letter we have extended the approach of  $I$ 
to  the calculation of the high  missing momentum part
of the processes $^3He(e,e^\prime p)^2H$ and  $^3He(e,e'p)(pn)$. We have found 
that our  theoretical predictions  provide
a very satisfactory agreement with  the Jlab data. Such an agreement 
cannot be considered fortuitous, for  it has been found to occur in 
the process $^2H(e,e'p)n$, as well as  in 
the processes $^3He(e,e^\prime p)^2H$ and  $^3He(e,e'p)(pn)$ in  kinematical conditions
differing  from the Jlab ones \cite{saclay} (see $I$).
A recent calculation \cite{laget2} of the same processes we have considered in this letter
also produced an overall good  agreement with the experimental data.
 The  calculation of \cite{laget2} is based upon the diagrammatic
expansion  developed in \cite{lagetgen}, and  the agreement with the experimental data 
at high values of the missing momentum is achieved by a three-body mechanism in which
the virtual photon is absorbed by a nucleon at rest which propagates emitting a meson,  which is
reabsorbed by the spectator nucleon pair. A careful comparison of  our results and the ones 
 of Ref.
\cite{laget2} reveals  however several  differences whose origin has to be clarified
in order to better understand  the basic
reaction mechanism of the process. To this end, it would be extremely useful to access:
 i)experimental
data at higher values of $p_m$ for the process $^3He(e,e'p)X$, and ii)
  experimental data for  the process
$^4He(e,e'p)^3H$, whose $p_m$ dependence, within the GEA,  looks different
 from the one of the process  $^3He(e,e'p)^2H$ 
\cite{chl}.

The authors are  indebted to  A. Kievsky  for making available
the variational three-body  wave functions of the Pisa Group.
Useful discussions  with M. Sargsian, R. Schiavilla and M. Strikman  are gratefully acknowledged. 
L.P.K. is   indebted to  the University of Perugia and INFN,
Sezione di Perugia, for a grant and for warm hospitality.


\begin{thebibliography}{99}
%
\bibitem{jlab1}
                  M. M.  Rvachev {\it et al}, \texttt{nucl-ex/0409005v3}.
%
\bibitem{jlab2}   
                  F. Benmokhtar {\it et al}, \texttt{nucl-ex/0408015v2}.
%
\bibitem{prelim}
                  A. Saha, M. Epstein, E. Voutier (spokespersons),
                  TJLAB Experiment E-89-044 and {\it Private Communications}
%
\bibitem{laget}
                  J.-M. Laget, \texttt{nucl-th/0410003}.
%
\bibitem{claleo}
                  C. Ciofi degli Atti, L. P. Kaptari, Phys. Rev.,\,\textbf{C71} (2005) 024005.
                  
%
\bibitem{rocco}
                  R. Schiavilla, {\it Private communication}.
%
\bibitem{pisa}
                  A. Kievsky, S. Rosati and M. Viviani,
                  Nucl. Phys. \textbf{A551} (1993) 241.
%
\bibitem{av18}
                  R. B. Wiringa, V. G. J. Stoks and  R. Schiavilla,
                  Phys. Rev. {\bf C51} (1995) 38.
%
\bibitem{mark}
                  L.L. Frankfurt, M.M. Sargsian, M.I. Strikman,
                  Phys. Rev. \textbf{C56} (1997) 1124.\\
                  M.M. Sargsian, T. V. Abrahamyan, M.I. Strikman, L.L. Frankfurt,
                  \texttt{nucl-th/0406020}.
%
\bibitem{glauber}
                  R. J. Glauber, in {\it Lectures in Theoretical Physics}, W. E. Brittin
                  {\it et al} Editors, New York (1959).

%
\bibitem{gribov}
                  V.N.Gribov, Sov. Phys. JETP, {\bf 30} (1970) 709.
%
\bibitem{bertocchi}
                  L. Bertocchi, Nuovo Cimento, {\bf 11A} (1972) 45.
%
\bibitem{nikolaev}
                  N. N. Nikolaev, J. Speth, B. G. Zakharov, J. Exp. Theor. Phys.
                  {\bf 82} (1996) 1046.
%
\bibitem{said}
                  R.A. Arndt {\it et al}, "(SAID) Partial-Wave Analysis Facility",
                  http://said.phys.vt.edu/.
%
\bibitem{forest}
                  T. de Forest Jr., Nucl. Phys. {\bf A392} (1983) 232.
%
\bibitem{elastic}
                  G. Alberi, G. Goggi, Phys. Reports, {\bf 74} (1981) 1.
%
\bibitem{saclay}
                  C. Marchand {\it et al},
                   Phys. Rev. Lett. \textbf{60} (1988) 1703.
%
\bibitem{laget2}  
                  J.-M. Laget, \texttt{nucl-th/0410003}
%
\bibitem{lagetgen}
                   J.-M. Laget, Phys. Rev. {\bf C38} (1988)2993.
%
\bibitem{chl}
                  C. Ciofi degli Atti, H. Morita, L. Kaptari, {\it to appear}.
%


\end{thebibliography}
\end{document}